**Rick Szostak** rszostak@ualberta.ca
**Andrea Scharnhorst** andrea.scharnhorst@dans.knaw.nl
**Wouter Beek** w.g.j.beek@vu.nl
**Richard P. Smiraglia** smiragli@uwm.edu

# Connecting KOSs and the LOD Cloud

**Abstract**
This paper describes a specific project, the current situation leading to it, its project design and first results. In particular, we will examine the terminology employed in the Linked Open Data cloud and compare this to the terminology employed in both the Universal Decimal Classification and the Basic Concepts Classification. We will explore whether these classifications can encourage greater consistency in LOD terminology. We thus hope to link the largely distinct scholarly literatures that address LOD and KOSs.

### 1.0 Introduction and Motivation

Our research[1] involves comparing the terminology employed within the Linked Open Data (LOD) Cloud with terminology employed within two KOSs: The Universal Decimal Classification (UDC) and the Basic Concepts Classification (BCC). In doing so we will connect two quite distinct literatures and communities of practice: the Semantic Web (SW) community, which has tended to be centered in computer science, and the knowledge organization (KO) community. In the SW community there have been increasing efforts to curate and preserve the machine-readable knowledge items as published on the Web using linked data formats (Beek, Rietveld at al. 2014; Beek at al. 2014). Controlled vocabularies play a prominent role in these efforts. They provide a way to index the knowledge graph, and they represent a semantically enriched layer in this graph. In knowledge organization (KO), systematic studies of KOSs have been proposed already (Tennis 2012), and such studies have also been executed for a number of small samples.

The promise of the web-based LOD Cloud is to free up data, metadata and information to a large extent from what often is called "data silos"—isolated information systems, which come with their own domain-specific knowledge organization systems, and are often barely interoperable. The LOD Cloud promises to deliver machine-readable KOSs and their implementation in a way that enables easy cross-linking. For example, the platform GeoNames (http://www.geonames.org) publishes about eleven billion place names in machine readable form, and has been used by many other services to relate a term like "New York" to a specific geographic reference, which in turn enables other services to link other names to this location, e.g., "City of New York," "New York City," or the historic term "Nieuw Amsterdam."

To be able to compare the different terminologies expressed in vocabularies, one first has to have an overview of them. Hence, our research involves the initial step of surveying the terminologies that are currently employed in linked open data. This will result in an atlas of vocabularies.

---

[1] Digging Into the Knowledge Graph, 2016 Digging Into Data Challenge
https://diggingintodata.org/awards/2016/project/digging-knowledge-graph

The SW holds the promise that different information repositories can all be encoded in the flexible graph-based representation language Resource Description Framework (RDF). Atomic statements in RDF take the form of triples, which are composed of a subject, predicate, and object term. RDF relies on URIs in order to assign universally unique names to concepts and instances. Since RDF names also denote locations on the internet, it uses RUIs for both naming and locating. By navigating to a URI location, software agents are able to extract the description of the entity denoted by that URI. Besides these syntactic and infrastructural properties, RDF also has a model-theoretic semantics that allows inferences to be drawn mechanically across different sets of information. If, for example, one website contains the RDF triple "Birds have wings" and another website contains the RDF triple "Penguins are birds," a computer can infer that "Penguins have wings." But this will only work if the same, or interoperable, terminology is employed. At present a wide variety of controlled vocabularies are employed across the LOD Cloud, but their formal semantics, including the inferences that follow from it, are not yet studied on a large web-scale.

Areas that make extensive use of SW technology include the humanities and the arts, as well as the social sciences (e.g., Hyvönen 2012). Mirroring the large variety of social and cultural phenomena in these fields, we find very specific, context-rich vocabularies developed by research communities as well as curators of collections. Increasingly, traditional curators of such vocabularies (e.g., The Getty Research Institute) provide their vocabularies as LOD.[2] In consequence, big data projects in the social sciences and humanities embrace SW technology (e.g., Hyvönen 2012). The ultimate goal of the collaboration in this project is to enhance the findability of facts and vocabulary used in the LOD Cloud and to enable scholars in the social sciences and humanities to find the right points to connect to when publishing LOD.

In the standardization of the SW (Horrocks 2003), a conscious choice was made to not create a specific upper ontology. Instead, SW standards like the Web Ontology Language (OWL) define a logical language that is devoid of content. Specific ontologies can be defined in terms of the logical primitives provided by these languages, but no specific conceptualization is prioritized. Specifically, the idea is that alternative, complementary, and even contradictory ontologies can be defined. This implements one of the core goals of the SW, which is to ensure that "anyone can say anything about anything."

In addition to the ability to encode multiple ontologies, the SW was designed in order to scale to facilitate the world-wide interchange of knowledge. Taking ideas from the World Wide Web (WWW), which is very successful in facilitating the world-wide interchange of documents, scalability is believed to be hampered by the existence of a centralized authority that coordinates the creation, distribution and retrieval of specific ontologies.

An inevitable result of these design goals is that there is currently not a complete overview of the ontologies that are published on the SW, although there are manually curated collections like Linked Open Vocabularies (LOV) (VanDenBussche 2017).

---

[2] See for example http://www.getty.edu/research/tools/vocabularies/lod/index.html

Even though there is a limited overview of which ontologies are present on the SW, datasets that are part of theLOD Cloud are known to often reuse a core set of popular existing vocabularies (Schmachtenberg 2014). In order to obtain a better overview of the ontologies and datasets that are currently published as LOD, the LOD Laundromat was developed at VU University Amsterdam.

But the collection of web-based vocabularies is only a first step. We will then proceed to compare the terminologiesthat are published in the LOD Cloud with the controlled vocabularies of the UDC and BCC. Note here that the challenge of interoperability across the LOD Cloud is itself a KO challenge; there is no explicit coordination between LD vocabularies.

There has been limited communication between the KO community and those active in developing the Semantic Web. We chose the UDC and BCC because these classifications have explicitly grappled with interdisciplinarity, and have pursued a faceted approach to classification (on BCC see Szostak 2013). The potential of the Semantic Web will best be realized if connections can be drawn across repositories. We thus wonder whether KOSs that strive to facilitate interdisciplinarity can play a key role in encouraging interoperability in the LOD cloud. Can the terminology employed in the LOD cloud be connected to KOS controlled vocabularies? Can the hierarchies and other relationships recognized within KOSs be used to structure terminology in the LOD cloud?

We will use those two generic classifications (UDC, BCC) as reference systems to develop generic principles of indexing. We will use high-level topical categorisations (similar to the UDC classes) and facets (similar to UDC auxiliaries such as place, time, peoples, forms, languages, etc.). We will contrast this with the phenomenon-based approach of the BCC, and ask questions of What (is studied)?, Why, Who, Where, and When? These categorisations will be tested in the archived version of the LOD Laundromat and eventually implemented in the open web services of the "living" LOD Laundromat. In particular, we will explore how general classification systems such as the UDC and BCC can be used to index linked data in a way that allows searching for concepts across domains, without becoming lost in the richness of the KOSs embedded in the LOD. In other words, we aim at a kind of union catalog for the LOD Laundromat snapshot, which will also be archived along with the LOD Laundromat data collection itself. One key question we hope to investigate is how interdisciplinarity is present and expressed or hidden and undiscovered.

At present, anyone wishing to code data for the SW has to choose among a bewildering array of sources of terminology. The choices made will determine to which other data repositories a computer can connect your data. Our research can potentially ease the choice facing those wishing to employ LOD and expand the degree of interoperability. We hope, in particular, to develop recommendations for LOD publication for communities in the social sciences and humanities (SSH), with emphasis on improving the re-use of existing vocabularies (among which we will encourage interoperability). We will identify, evaluate and index SSH-relevant vocabularies by mapping clusters of similar meaning onto KOSs.

Though implications for the SW are perhaps most obvious, our research also has important implications for KO. If KOSs can play a critical role in encouraging interoperability across the SW, then the field of KO gains an important new audience for its work. Note that the premise of the SW is that data of all types need to be explicitly encoded in terms of formal languages such as RDF and OWL. In other words, the SW is grounded in the recognition that there are limits to what can be discovered by simply searching un- or weakly-structured natural language texts. The KO community's longstanding efforts to develop structured controlled vocabulary at times seems to be overshadowed by search algorithms that search natural language texts rather than structured metadata, but the SW potentially places KO at the center of future developments in search engines. In addition, research has shown that classifications themselves form navigable knowledge networks among the resources to which they are linked (Suchecki et al. 2012; Smiraglia et al. 2013).

Much effort is undertaken in the KO domain to bring KOSs into use in the LOD Cloud (e.g., Baca and Gill 2015). As mentioned above, there is effort to link general controlled vocabularies, such as The Getty Vocabularies, to the LOD Cloud. There are definite advantages in vocabulary mapping for people-centered properties (what librarians call "authority control" of names), for LOD, to alleviate the problems of property proliferation in LOD environments. The discourse concerning the SW reveals a research agenda for KOSs including direct linkage of domain-centric ontologies within the LOD Cloud, including most importantly for this project, vocabulary alignment. We hope to provide advice on how KOSs might be revised to reflect and serve the LOD Cloud (especially from the perspective of interdisciplinarity).

It should be stressed that the proposed research will provide a much-needed link between LOD and KOSs. By mapping one onto the other we can compare the structure of the two. KOSs always employ some sort of logical structure with enumerative capability: the idea that a place must be found in the KOS for all works or ideas. In addition, the traditional use of "literary warrant" allows a KOS to point directly to the documentary source of a particular concept. Comparing LOD clusters with a KOS can indicate where a particular KOS needs to be amended. That is, LOD clusters provide literary warrant for extending enumeration and clarifying the KOS. In turn, the mapping can suggest how LOD can be better structured/indexed to facilitate the practice of actually linking data. We can thus harness the wisdom of the KO community to the important practice of achieving interoperability or even consensus on LOD terminology. To achieve this interplay with respect to the BCC we must render the BCC into LOD terminology and then compare the result with the clusters of LOD terminology we obtain.

The next three short segments of this paper describe deliverables from our research as of January 2018, midway through the first year of a three year project.

**2.0 Toward a KOS Observatory**

A challenge frequently articulated across the KO and SW communities is the ability to track and maintain access to changing KOSs across time and across applications. In addition to Beek et al. (2014), Tennis (2002; 2007; 2012; 2015; 2016) has been the most prominent catalyst for the KO community. Two KNOWeSCAPE workshops were held in Amsterdam (2015) and Malta (2017) to bring together experts from the KO, SW, publishing and digital humanities communities to prioritize objectives for visioning and creating an observatory for KOSs.[3] Following on these workshops and in conjunction with our research, an effort to map a small initial set of KOSs has been undertaken by the DANS (Data Archiving and Networked Services) division of the Royal Netherlands Academy of the Arts and Sciences (KNAW). Our team is working on a project of larger scope than the similar Basel Register of Thesauri, Ontologies & Classifications (BARTOC, https://bartoc.org), but consulting BARTOC along with standard bibliographic and internet resources we have created a template and begun building a database of KOSs. Figures 1a-b show parts of the experimental template for *Art & Architecture Thesaurus*.

*Figure 1a*. Observatory template for Art & Architecture Thesaurus, part 1, edited.

| Creator(s)/Curator(s) | Maintenance organization | Format(s) | Physical Location | Online Location |
|---|---|---|---|---|
| Toni Petersen | Getty Art History Information Program | Printed Book. | KOS.2!A1:A13 | n/a |
| Toni Petersen | Getty Art History Information Program | Printed Book. | KOS.2!A26:A27 | n/a |
| Toni Petersen | Getty Art History Information Program | Printed Book, eBook, Online, Data Files, Computer Discs. | KOS.2!A15 | KOS.2!B15:B23 |
| Toni Petersen | Getty Art History Information Program | Digital, 6 computer discs ; 3 1/2-5 1/4 in. + 1 Introduction to the art and architecture thesaurus, 2nd ed. (250 pages ; 27 cm) + 1 quick-reference card + 1 user's manual for the Authority references tool ([126] pages ; 27 cm) | KOS.2!A26:A27 | n/a |
| Toni Petersen | Getty Art History Information Program | Digital, 3 computer discs ; 3 1/2 in. + 2 user manuals (27 cm) + 1 quick reference card (27 x 57 cm folded to 27 x 19 cm) + 1 general information card (24 x 57 cm folded to 24 x 15 cm) + 1 demo disc | n/a | n/a |

*Figure 1b*. Observatory template for Art & Architecture Thesaurus, part 2, versioning data, edited.

| Identifier | Schema Name/Title | Earlier versions (editions) … | History of versioning: |
|---|---|---|---|
| KOS.2 | Art & Architecture Thesaurus; AAT | n/a | 1st Edition |
| KOS.2.1 | Art & Architecture Thesaurus; AAT | 1st edition 1990 | 1st Edition: Supplement 1 |
| KOS.2.2 | Art & Architecture Thesaurus; AAT | 1st edition supplement 1, 1992 | 2nd Edition |
| KOS.2.3 | Art & Architecture Thesaurus; AAT | 2nd edition 1994 | 2nd Edition: Version 2.0 |
| KOS.2.4 | Art & Architecture Thesaurus; AAT | 2nd edition 1994 | 2nd Edition: Version 2.1 |

---

[3] Both workshops were sponsored by COST Action grant TD1210. See Evolution and variation of classification systems – KnoweScape workshop March 4-5, 2015 Amsterdam (http://knowescape.org/evolution-and-variation-of-classification-systems-knowescape-workshop-march-4-5-2015-amsterdam/) and Workshop Observatory for Knowledge Organisation Systems in Malta, Feb 1-3, 2017 (http://knowescape.org/event/observatory-knowledge-organisation-systems/).

| KOS.2.5 | Art & Architecture Thesaurus; AAT | 2nd edition 1994 | Online |
| KOS.2.7 | AAT-Deutsch | Online | Online, Version DE |
| KOS.2.8 | TAA; El Tesauro de Arte & Arquitectura | Online | Online, Version ES |

Because of space limitations these illustrations have been edited. By experimental template, we mean that during the course of collecting information about KOSs used in the social sciences and humanities domain, our own KO–as embodied in the template is changing, depending on the information available on the KOSs, and our growing insights.

However, it is clear to see that because of our aim to link these KOSs to the LOD Cloud as well as to enable version tracking we are collecting detailed data on location and versions for each system. It is our plan to archive our own data in the DANS EASY online archiving system for research data (https://easy.dans.knaw.nl/ui/home).

**3.0 Classification as LOD**

A major goal of the first year of research is to create both generic classifications (UDC and BCC) as linked data. This research is progressing in two separate streams. BCC is being rendered as LOD by the University of Alberta project team. The UDC is a proprietary system, however, so mixed solutions will be required. The UDC Summaries online, which are publicly available via the internet already exist as LOD (http://udcdata.info/). It is important that the entire UDC be rendered as linked data, but the proprietary portions may remain behind paywalls for use by licensed subscribers.

**4.0 Classifying concepts (not documents)**

Classifying concepts is different from classifying documents. Typical document classification is tied to a summary of the overall subject of a document, in order to place the document among related texts. Our goal, however, is to use both classifications to point to specific concepts, rather in the style of the famous grinder metaphor of Paul Otlet (see Figure 1 in Smiraglia and van den Heuvel 2013, 363). In other words, we hope to create networked linkages in the LOD cloud among concepts, their classification nodes, and the SW resources to which each are linked. Research comparing the use of the two classifications empirically in a set of OCLC WorldCat data has revealed the problem of concepts that are less explicit in UDC because they are hidden hierarchically (Szostak and Smiraglia 2018). The phenomenon-based BCC is more efficacious for directly pointing to concepts. The two classifications together, however, provide the best of both worlds by pointing directly to concepts and also placing them in disciplinary contexts.

**5.0. Conclusion**

The encounter between SW and KO comes naturally with the need to sort out terminology, concepts, and epistemic values. One task of this project, for which this paper is an example, is to provide "translations" from one knowledge domain into the other, to create a commonly shared understanding and enabling communication; in short, to establish a trading zone as science and technology studies would call this (Galison 1997). Another precondition to make specific KOSs operable in the LOD cloud is their publication as LOD. This task can rely on some traditions of the adoption of LD principles for those curating KO. Much more challenging is the task to explore to which extent those new generic KOSs can be made of use to a) increase findability of KOSs in the LOD; and, b) to connect UDC and BCC to other parts of the LOD so that their power of expressing concepts and describing phenomena in a manner that is universal, cross-cultural and cross-domain can be turned into a navigation tool inside of the LOD.

**Acknowledgment**
We would like to acknowledge Gerard Coen, DANS, for contributions to the initial mapping of KOSs including those shown here in Figures 1a-b.

**References**
Baca, Murtha and Melissa Gill. 2015. "Encoding Multilingual Knowledge Systems in the Digital Age: the Getty Vocabularies." *Knowledge Organization* 42: 232-43.
Beek, Wouter, Laurens Rietveld, Hamid R. Bazoobandi, Jan Wielemaker and Stefan Schlobach. 2014. "LOD Laundromat: A Uniform Way of Publishing Other People's Dirty Data." *The Semantic Web, ISWC 2014: 13th International Semantic Web Conference, Riva del Garda, Italy, October 19-23, 2014. Proceedings, part 1*, ed. Peter Mika et al. Lecture Notes in Computer Science 8796. Cham: Springer, 213-28. doi:10.1007/978-3-319-11964-9_14
Beek, Wouter, Paul Groth, Stefan Schlobach and Rinke Hoekstra. 2014. "A Web Observatory for the Machine Processability of Structured Data on the Web." *WebSci '14: Proceedings of the 2014 ACM Conference on Web Science, Bloomington, Indiana, USA, June 23-24, 2014*. New York: ACM, 249-50. doi:10.1145/2615569.2615654
Galison, Peter. 1997. *Image and Logic: A Material Culture of Microphysics*. Chicago: University of Chicago Press.
Horrocks, Ian, Peter F. Patel-Schneider and Frank Van Harmelen. 2003. "From SHIQ and RDF to OWL: The Making of a Web Ontology Language." *Web Semantics: Science, Services and Agents on the World Wide Web* 1, no. 1: 7-26.
Hyvönen, Eero. 2012. *Publishing and Using Cultural Heritage Linked Data on the Semantic Web*. Synthesis Lectures on the Semantic Web: Theory and Technology 2. [San Rafael]: Morgan and Claypool. doi:10.2200/S00452ED1V01Y201210WBE003
Klein, Martin, Herbert Van de Sompel, Robert Sanderson, Harihar Shankar, Lyudmila Balakireva, Ke Zhou and Richard Tobin. 2014. "Scholarly Context Not Found: One in Five Articles Suffers from Reference Rot." *PLoS ONE* 9, no. 12: e115253. doi:10.1371/journal.pone.0115253
Koehler, Wallace. 2002. "Web Page Change and Persistence: A Four-year Longitudinal Study." *Journal of the American Society for Information Science and Technology* 53: 162–171. doi:10.1002/asi.10018
Schmachtenberg, Max, Christian Bizer and Heiko Paulheim. 2014. "Adoption of the Linked Data Best Practices in Different Topical Domains." In *International Semantic Web Conference*. Springer, Cham, 245-60.

Smiraglia, Richard P. and Charles van den Heuvel. 2013. "Classifications and Concepts: Towards an Elementary Theory of Knowledge Interaction." *Journal of Documentation* 69: 360-83.

Smiraglia, Richard P., Andrea Scharnhorst, Almila Akdag Salah and Cheng Gao. 2013. "UDC in Action." In *Classification and Visualization: Interfaces to Knowledge, Proceedings of the International UDC Seminar, 24-25 October 2013, The Hague, The Netherlands*, ed. Aïda Slavic, Almila Akdag Slah and Sylvie Davies eds. Würzburg: Ergon, 259-72.

Suchecki, Krzysztof, Alkim Almila Akdag Salah, Cheng Gao and Andrea Scharnhorst. 2012. "Evolution of Wikipedia's Category Structure." *Advances in Complex Systems* 15 supp01: 1250068.

Szostak, Rick. 2013. *Basic Concepts Classification*. https://sites.google.com/a/ualberta.ca/rick-szostak/research/basic-concepts-classification-web-version-2013

Smiraglia, Richard P. and Rick Szostak. 2018. "Converting UDC to BCC: Comparative Approaches to Interdisciplinarity." In *Proceedings of the Fifteenth International ISKO Conference, Porto, Portugal, July 9-11, 2018*. Würzburg: Ergon, forthcoming.

Tennis, Joseph T. 2002. "Subject Ontogeny: Subject Access through Time and the Dimensionality of Classification." In *Challenges in Knowledge Representation and Organization for the 21st Century: Integration of Knowledge across Boundaries: Proceedings of the Seventh International ISKO Conference, 10-13 July 2002, Granada, Spain*, ed. M. J. López-Huertas. Advances in knowledge organization 8. Würzburg: Ergon Verlag, 54-9.

Tennis, Joseph T. 2007. "Scheme Versioning in the Semantic Web." *Cataloging & Classification Quarterly* 43, no.3: 85-104. doi: 10.1300/J104v43n03

Tennis, Joseph T. 2012. "The Strange Case of Eugenics: A Subject's Ontogeny in a Long-lived Classification Scheme and the Question of Collocative Integrity." *Journal of the American Society for Information Science and Technology* 63: 1350-59. doi:10.1002/asi.22686

Tennis, Joseph T. 2015. "The Memory of What Is: Ontogenic Analysis and its Relationship to Ontological Concerns in Knowledge Organization." In *Ontology for Knowledge Organization*, ed. Richard P. Smiraglia and Hur-Li Lee. Würzburg: Ergon-Verlag, 97-100.

Tennis, Joseph T. 2016. "Methodological Challenges in Scheme Versioning and Subject Ontogeny Research." *Knowledge Organization* 43: 573-80.

Vandenbussche, Pierre-Yves, Ghislain A. Atemezing, María Poveda-Villalón and Bernard Vatant. 2017. "Linked Open Vocabularies (LOV): A Gateway to Reusable Semantic Vocabularies on the Web." *Semantic Web* 8: 437-52.